\documentstyle[11pt,epsfig]{article}

\newcommand{\be}{\begin{equation}}
\newcommand{\ee}{\end{equation}}
\newcommand{\ba}{\begin{eqnarray}}
\newcommand{\ea}{\end{eqnarray}}

[1999/03/29 PSNFSS v.7.2
Times font as default roman
: S Rahtz]

\begin{document}
\setlength{\baselineskip}{.5cm}
\renewcommand{\thefootnote}{\fnsymbol{footnote}}
\newcommand{\lp}{\left(}
\newcommand{\rp}{\right)}

\begin{center}
\centering{\bf \Large Predictability of catastrophic events: \\
material rupture, earthquakes, turbulence, \\financial crashes and human birth\\
{\normalsize Didier Sornette}}
\end{center}
\begin{center}
\centering{
Institute of Geophysics and Planetary Physics\\ and
Department of Earth and Space Sciences\\
UCLA, Box 951567, Los Angeles, CA 90095-1567\\
and Laboratoire de Physique de la Mati\`ere Condens\'ee, CNRS UMR6622\\
Universite de Nice-Sophia Antipolis, Faculte des Sciences\\
06108 NICE Cedex 2, France\\
}
\end{center}

\bigskip

\bigskip

\begin{abstract}

We propose that catastrophic events are ``outliers'' with statistically different
properties than the rest of the population and result from mechanisms involving
amplifying critical cascades.
Applications and the potential for prediction are discussed
in relation to the rupture of composite materials, great earthquakes, 
turbulence and abrupt changes of weather regimes, financial crashes
and human parturition (birth).

\end{abstract}





\vskip 1cm

\section{Introduction}

What do a high pressure tank on a rocket, a seismic fault 
and a busy market have in common? 
Recent research suggests that they can all be described in much the
same basic physical terms\,: as self-organising systems which develop similar patterns
over many scales, from the very small to the very large. And all three have the
potential for extreme behaviour\,: rupture, quake or crash.

Similar characteristics are exhibited by other crises that often present
fundamental societal 
impacts and range from large natural catastrophes such as volcanic
eruptions, hurricanes and tornadoes, landslides, avalanches, lightning strikes,
catastrophic events of environmental degradation,
to the failure of engineering structures, social
unrest leading to large-scale strikes and upheaval, economic drawdowns on national
and global scales, regional power blackouts, traffic gridlock, diseases and
epidemics, etc. 
Intense attention and efforts are devoted in the academic community,
in goverment agencies and in the industries that are sensitive to or directly
interested in these risks, to the understanding, assessment,
mitigation and if possible prediction of these events. 

A central property of such complex systems is the possible occurrence of
coherent large-scale collective behaviors with a very rich structure, resulting from the 
repeated non-linear interactions among its constituents: the whole turns out 
to be much more than the sum of its parts. It is widely believed that
most complex systems are not amenable to mathematical, analytic descriptions
and can only be explored by means of ``numerical experiments''. In the context
of the mathematics of algorithmic complexity \cite{algocompl}, many complex systems are 
said to be computationally irreducible,
i.e. the only way to decide about their evolution is to actually let them evolve in
time. Accordingly, the ``dynamical'' future time evolution of 
complex systems would be inherently unpredictable. This unpredictability does not
prevent however the application of the scientific method for the prediction of novel phenomena
as exemplified by many famous cases (prediction of the planet Neptune by Leverrier from
calculations of perturbations in the orbit of Uranus, the prediction
by Einstein of the deviation of light by the sun's gravitation field, 
the prediction of the
helical structure of the DNA molecule by Watson and Crick based on earlier predictions
by Pauling and Bragg, etc.).
In contrast, it refers to the frustration to satisfy the quest for the knowledge of
what tomorrow will be made of, often filled by the vision of 
 ``prophets'' who have historically inspired or terrified the masses.
 
The view that complex systems are inherently unpredictable has
recently been defended persuasively in concrete prediction applications, such as 
the socially important issue of earthquake prediction (see the 
contributions in \cite{Naturedebate}). In addition to the persistent 
failures to find a reliable earthquake predictive scheme, this view is rooted 
theoretically in the analogy between earthquakes and self-organized criticality 
\cite{Bak,mybook}.
In this ``fractal'' framework, there is no characteristic scale and the power law
distribution of earthquake sizes reflects the fact that 
large earthquakes are nothing but small earthquakes that did not stop. They
are thus unpredictable because their nucleation is not different from that of
the multitude of small earthquakes which obviously cannot be all predicted
\cite{Geller}.

Is it really impossible to predict
all features of complex systems? Take our personal life.
We are not really interested in knowing in 
advance at what time we will go to a given store or drive on a highway. We are much more 
interested in forecasting the major bifurcations ahead of us, involving the 
few important things, like health, love and work that count for our happiness.
Similarly, predicting the detailed evolution of complex systems has no real value and the
fact that we are taught that it is out of reach from a fundamental point of view does
not exclude the more interesting possibility of predicting phases of evolutions of 
complex systems that really count, like the extreme events. 

It turns out that 
most complex systems in the natural and social sciences do
exhibit rare and sudden transitions, that occur over
time intervals that are short compared to the
characteristic time scales of their posterior evolution.
Such extreme events express more than anything else the underlying ``forces'', usually
hidden by an almost perfect balance and thus provide the potential for a better
scientific understanding of complex systems. 

It is essential to realize that
the long-term behavior of these complex systems is
often controlled in large part by these rare catastrophic events: the universe
was probably born during an extreme explosion (the ``big-bang''); the nucleosynthesis of
all important heavy atomic elements constituting our matter results from the colossal
explosion of supernovae (these stars heavier than our sun whose
internal nuclear combustion diverges at the end of their life); the largest
earthquake in California repeating about once every two centuries accounts for
a significant fraction of the total tectonic deformation; landscapes are more
shaped by the ``millenium'' flood that moves large boulders rather than the action
of all other eroding agents; the largest volcanic eruptions lead to major 
topographic changes as well as severe climatic disruptions; 
according to some contemporary views,  evolution is 
probably characterized by phases
of quasi-stasis interrupted by episodic bursts of activity and destruction
\cite{Gouldpunc1,Gouldpunc}; 
financial crashes, which can destroy in an instant trillions
of dollars, loom over and shape the psychological state of investors;
political crises and revolutions shape the long-term geopolitical 
landscape; even our personal life is shaped in the long run by a few key
decisions or events. 

The outstanding scientific question that needs to be addressed 
in order to guide prediction is
how large-scale patterns of a catastrophic
nature might evolve from a series of interactions on the smallest and increasingly
larger scales, where the rules for the interactions are presumed identifiable and known.
For instance, a typical report on an industrial catastrophe describes the improbable
interplay between a succession of events. Each event has a small probability and limited
impact in itself. However, their juxtaposition and chaining lead inexorably to the
observed losses. A common denominator of the various examples of crises is that they
emerge from a collective process: the repetitive actions of interactive nonlinear
influences on many scales lead to a progressive build-up of large-scale correlations
and ultimately to the crisis. In such systems, it has been found that the organization 
of spatial and temporal correlations does not stem, in general, from a nucleation phase
diffusing accross the system. It results rather from a progressive and more global
cooperative process occurring over the whole system by repetitive interactions. 

For hundreds of years, science has proceeded on the notion that things can
always be understood--and can only be understood--by breaking them down into smaller
pieces, and by coming to know these pieces completely. Systems in critical
states flout this principle. Important aspects of their
behaviour cannot be captured knowing only the detailed properties of their component parts.
The large scale behavior is controlled by their cooperativity and scaling up of their
interactions. This is the key idea underlying the 
four examples that illustrate this new approach to prediction\,: rupture of
engineering structures, earthquakes, stock market crashes and human parturition (birth).

\section{Prediction of rupture in complex systems}

\subsection{Nature of the problem}

The damage and fracture of materials are technologically of major
interest because of their economic and human cost. They cover a wide range of 
phenomena such as cracking of glass, aging of concrete, the failure of fiber
networks, and the breaking of a metal bar subject to
an external load. Failures of composite systems are of upmost importance in
the naval, aeronautics and space industries. By the term composite, we
include both materials with constrasted microscopic structures and assemblages
of macroscopic elements forming a super-structure. Chemical and nuclear
plants suffer from cracking due to corrosion either of chemical or radioactive
origin, aided by thermal and/or mechanical stress. More exotic but no less
interesting phenomena include the fracture of an old painting, the
pattern formation of the cracks of drying mud in deserts, and rupture
propagation in earthquake faults. 

Despite the large amount of
experimental data and the considerable effort that has been undertaken by
material scientists, many questions about fracture and fatigue have not yet
been answered. There is no comprehensive understanding of rupture
phenomena, but only a partial classification in
restricted and relatively simple situations. This lack of fundamental
understanding is reflected in the absence of proper prediction methods for
rupture and fatigue, that can be based on a suitable monitoring of the stressed system.

\subsection{The role of heterogeneity}

In the early sixties, the Japanese seismologist K. Mogi \cite{Mogi1,Mogi2}
noticed that the fracture process depends strongly on the degree of
heterogeneity of materials\,: the more heterogeneous, the more warnings
one gets\,; the more perfect, the more treacherous is the rupture. The
failure of perfect crystals thus appears to be unpredictable while the fracture of
dirty and deteriorated materials may be forecast. For once, complex
systems could be simpler to comprehend! However, since its
inception, this idea has not been developed because it is hard to
quantify the degrees of ``useful'' heterogeneity, which probably depend
on other factors such as the nature of the stress field, the presence of
water, etc. In our work on the failure of mechanical systems, 
we have solved this paradox quantitatively using
concepts inspired from statistical physics, a domain where complexity has
long been studied as resulting from collective behavior. The idea is that,
upon loading a heterogeneous material, single isolated microcracks appear
and then, with the increase of load or time of loading, they both grow and
multiply leading to an increase in the number of cracks.
As a consequence, microcracks begin to merge until a ``critical density''
of cracks is reached at which time the main fracture is formed. It is
then expected that various physical quantities (acoustic emission, elastic,
transport, electric properties, etc.) will vary. However, the nature of
this variation depends on the heterogeneity. The new result is that there
is a threshold that can be calculated\,: if disorder is too small, then the
precursory signals are essentially absent and prediction is impossible. 
If heterogeneity is large, rupture is more continuous.

To obtain this insight, we used simple mechanical models of masses and springs
with local stress transfer \cite{Andersen1}. This class of models provides
a simplified description in order to identify the different
regimes of behavior. The scientific enterprise is paved with such reductionism
that has worked surprisingly well. We were thus able to quantify how
heterogeneity plays the role of a relevant field\,: systems with limited
stress amplification exhibit a so-called tri-critical transition,
from a Griffith-type abrupt rupture (first-order) regime to a
progressive damage (critical) regime as the disorder
increases. This effect was also demonstrated on a
simple mean-field model of rupture, known as the
democratic fiber bundle model. 
 It is remarkable that the disorder is so relevant as to change
the nature of rupture. In systems with long-range elasticity, 
the nature of the rupture process may not
change qualitatively as above, but quantitatively\,: any disorder may be relevant
in this case and make the rupture similar to a critical point; however, we have recently
shown that the disorder controls the width of the critical region \cite{Andersen2}. The
smaller it is, the smaller will be the critical region, which
may become too small to play any role in practice. For
realistic systems, long-range correlations transported by the stress field
around defects and cracks make the problem more subtle. Time dependence is
expected to be a crucial aspect in the process of correlation building in these
processes. As the damage increases, a new ``phase'' appears, where micro-cracks 
begin to merge leading to screening and other cooperative effects. 
Finally, the main fracture is formed leading to global failure. 
In simple intuitive terms, the failure of compositive systems
may often be viewed as the result of a correlated 
percolation process. The challenge is to describe the transition from the
damage and corrosion processes at the microscopic level to the macroscopic
rupture.

\subsection{Scaling, critical point and rupture prediction}

Motivated by the multi-scale nature of the second class of ruptures and
analogies with the percolation model, physicists
working in statistical physics started to suggest
in the mid-eighties 
that rupture of sufficiently heterogeneous media would exhibit some
universal properties, in a way similar to critical phase transitions. The idea
was to build on the knowledge accumulated in statistical physics on
the so-called $N-$body problem and cooperative effects in order to describe
multiple interactions between defects. However, 
most of the models were quasi-static
with very simple loading rules.
Scaling laws were found to describe size effects and damage properties.

In 1992, we proposed the first model of rupture with a realistic dynamical law
for the evolution of damage, modelled as a space dependent damage variable, a realistic
loading and with many growing interacting micro-cracks \cite{Vanneste1,Vanneste2,Vanneste3}. 
We found that the total rate of damage, as measured for instance by
the elastic energy released per unit time, increases as a power law of the 
time-to-failure on the approach to the global failure.
In this model, rupture was indeed found 
to occur as the culmination of the progressive nucleation, growth and fusion
between microcracks, leading to a fractal network, but the exponents 
were found to be non-universal and a function of the damage law. This
model has since been found to describe correctly experiments on 
the electric breakdown of insulator-conducting composites \cite{Lamai}.
Another application is damage
by electromigration of polycristalline metal films \cite{Bradley1,Bradley2}.

In 1993, we extended these results by testing
on engineering composite structures the
concept that failure in fiber composites may be described by a critical state,
thus predicting that the rate of damage would exhibit a power law behavior
\cite{Anifrani}. This critical behavior may correspond to an acceleration of the 
rate of energy release or to a deceleration, depending on the nature and range of the 
stress transfer mechanism and on the loading procedure. We based our
approach on a theory of 
many interacting elements called the renormalization group. The renormalization group 
can be thought of as a construction scheme or ``bottom-up''
approach to the design of large scale structures. 
Since then, other numerical simulations of 
statistical rupture models and controlled experiments have
confirmed that, near the global failure point, 
the cumulative elastic energy released during fracturing of heterogeneous
solids follows a power law behavior.

Based on an extension of the usual solutions of the renormalization group
and on explicit numerical and theoretical calculations, 
we were thus led to propose that the power law behavior of the time-to-failure
analysis should be corrected for the presence of log-periodic modulations \cite{Anifrani}.
Since then, this method has been tested extensively during our continuing collaboration
with the French Aerospace company A\'erospatiale
on pressure tanks made of kevlar-matrix
and carbon-matrix composites embarked on the European Ariane 4 and 5 rockets. 
In a nutshell, the method consists in
this application in 
recording acoustic emissions under constant stress rate
and the acoustic emission energy as a function of
stress is fitted by the above log-periodic critical theory. One of the parameters is the
time of failure and the fit thus provides a ``prediction'' when the sample
is not brought to failure in the first test. Improvements 
of the theory and of the fitting formula were applied to 
about 50 pressure-tanks. The results indicate that a precision of  
a few percent in the determination of the stress at rupture 
is obtained using acoustic emission recorded $20~\%$ below
the stress at rupture.
These successes have warranted an international patent and the selection of this non-destructive
evalution technique as the routine qualifying procedure in the industrial fabrication
process. See \cite{JohansenSorcritrup} for a recent synthesis and reanalysis.

This example constitutes a remarkable case where rather abstract theoretical concepts
borrowed from the rather esoteric field of statistical and nonlinear physics
 have been applied directly to a concrete industrial problem. 
This example is remarkable for another reason that we would like to describe now.

\subsection{Discrete scale invariance, complex exponents and log-periodicity}

During our research on the 
acoustic emissions of the industrial pressure tank of the European Ariane rocket,
we discovered the existence of log-periodic scaling in non-hierarchical systems.
 To fix ideas, consider the acoustic energy  
$E \sim (t_c - t)^{-\alpha}$ following a power law
 a function of time to failure. Suppose that there
is in addition a log-periodic signal modulation
\be
E \sim (t_c - t)^{-\alpha}~[1+ C\cos[2\pi {\log (t_c - t) \over \log \lambda}]]~.
\label{mqmmqm}
\ee
We see that the local maxima of the signal occur at $t_n$ such that
the argument of the cosine is a multiple to $2\pi$, leading to a geometrical time
series $t_c - t_n \sim \lambda^{-n}$ where $n$ is an integer. The 
oscillations are thus modulated in frequency with a 
geometric increase of the frequency on the  approach to the critical point $t_c$.
This apparent esoteric property turns out to be
surprisingly general both experimentally and theoretically and we are probably only at 
the beginning of our understanding. From a formal point of view,
log-periodicity can be shown to be nothing but the concrete expression 
of the fact that exponents or
more generally dimensions can be ``complex'', i.e. belong to these numbers which when 
squared give negative values.

\begin{figure}
\parbox[l]{7.5cm}{
\epsfig{file=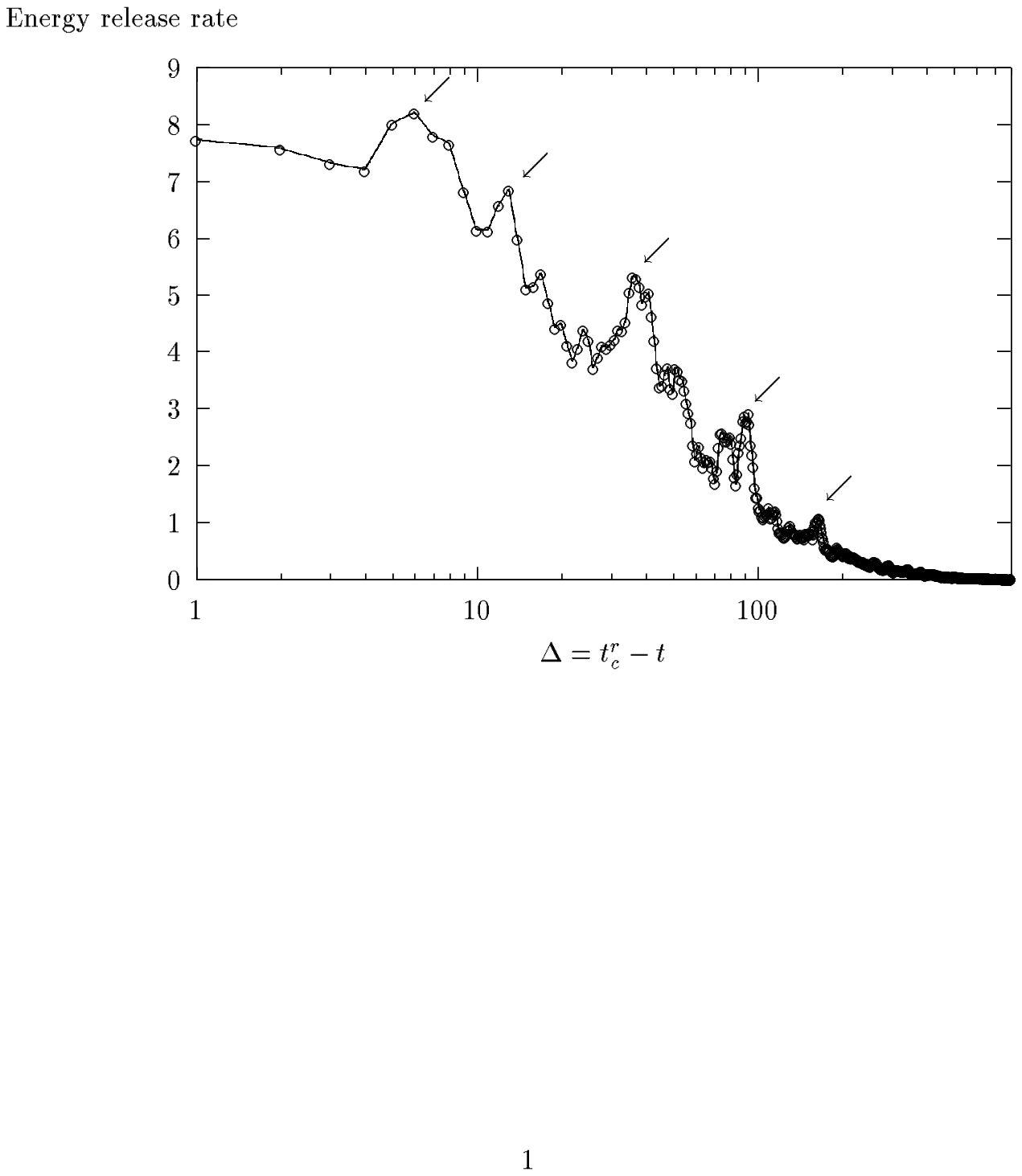,height=4cm,width=7cm,bbllx=150,bblly=300,bburx=500,bbury=500}
\caption{\label{sorjoh6} Log-log plot of the energy release rate of a
mechanical system approaching rupture. A so-called ``canonical'' averaging procedure
\cite{Canonicalproc}
has been performed over 19 independent systems to exhibit the characteristic time scales shown as
the arrows decorating the overall power law time-to-failure behavior.} }
\hspace{5mm}
\parbox[r]{7.5cm}{
\epsfig{file=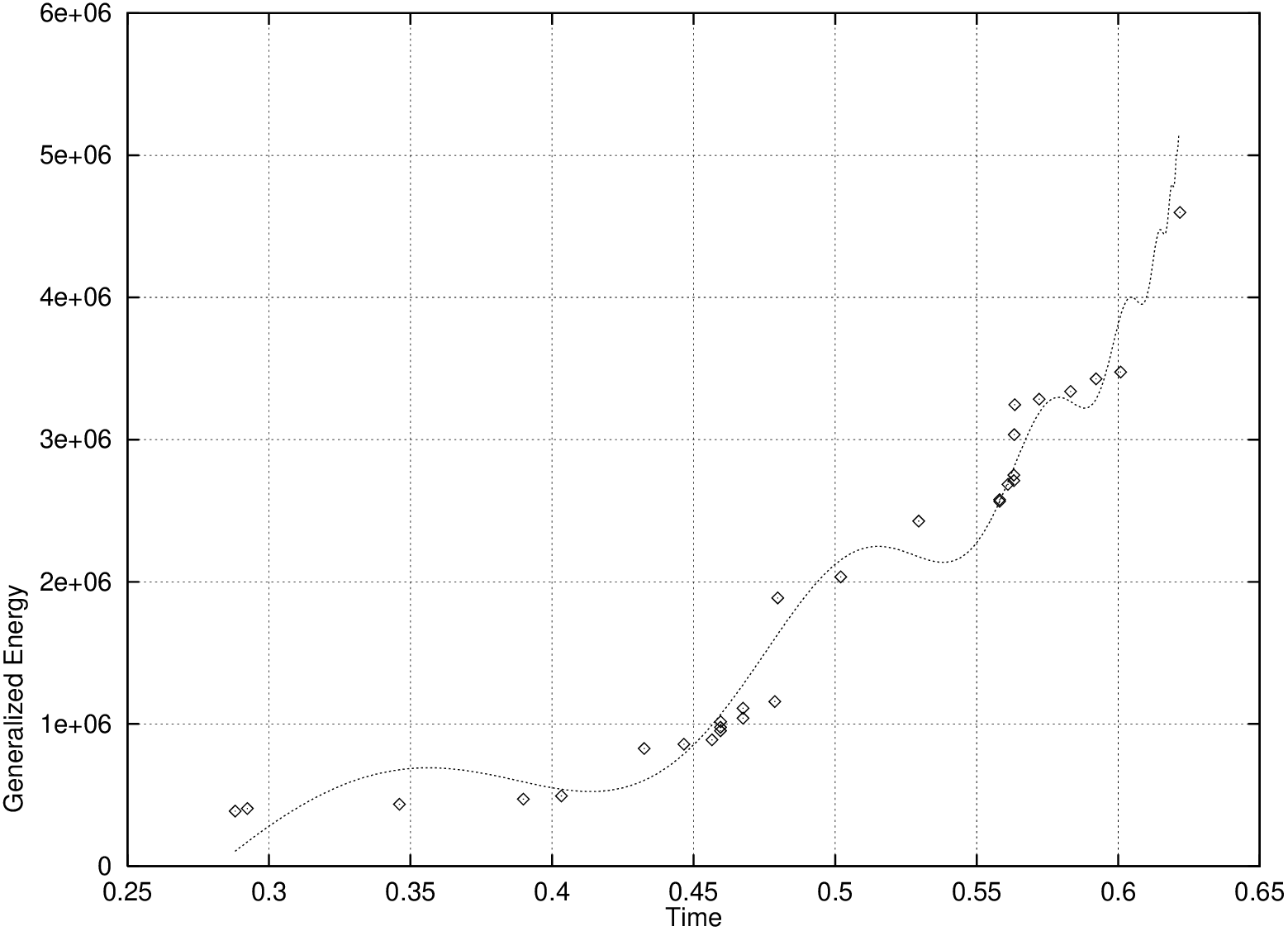,height=4cm,width=6cm,bbllx=100,bblly=-480,bburx=620,bbury=50}
\caption{\label{ISSI} Fit with eq.\ref{mqmmqm}, where $x=t_c-t$ and $t_c$
is the critical rupture time, of a cumulative seismic deformation rate
prior to a large rockburst in a deep South African mine. Such fits are also
used in a forecasting mode \cite{Ouillonmine}.} }
\end{figure}

In Fig.\ref{sorjoh6}, we illustrate how log-periodicity helps us to understand
the time dependence of
signals precursors to global failure in a numerical network of discrete
fuses. This so-called time-to-failure analysis is based on the
detection of an acceleration of some measured signal, for instance acoustic
emissions, on the approach to the global failure. 
Fig.\ref{ISSI} shows another application using (\ref{mqmmqm}) to fit
 the cumulative energy released by small earthquakes prior to
a major rockburst in a deep South African mine, where the data are of
particularly good quality \cite{Ouillonmine}. 

Encouraged by our observation of log-periodicity in rupture phenomena, we
started to investigate whether similar signatures could be observed in other systems.
And looking more closely, we were led to find them in many systems in which they 
had been previously unsuspected. These structures have long been known as possible 
from the formal solutions of renormalization group
equations in the seventies but were rejected as physically irrelevant.
They were studied in the eighties in a
rather academic context of special artificial hierarchical 
geometrical systems. Our work
led us to realize that discrete scale invariance and its associated complex
exponents and log-periodicity may appear ``spontaneously'' in natural systems, 
i.e. without the need for
a pre-existing hierarchy. Examples that we have documented \cite{logperiodicreport} 
are diffusion-limited-aggregation clusters, rupture
in heterogeneous systems, earthquakes, animals (a generalization of percolation) among
many other systems. Complex scaling could also be relevant to turbulence, to
the physics of disordered systems, as well as to the description of
 out-of-equilibrium dynamical systems.
Some of the physical mechanisms at the origin of these structures are now
better understood. General considerations using the framework of field
theories, the framework to describe fundamental particle physics
and condensed matter systems, show that they should constitute the rule rather than 
the exception, similarly to the realization that chaotic (non-integrable) dynamical
systems are more general that regular (integrable) ones. In addition to a fascinating
physical relevance of this abstract notion of complex dimensions, the even more 
important aspect in our point of view is that 
 discrete scale invariance and its signatures may provide
new insights in the underlying mechanisms of scale invariance and be very useful
for prediction purposes.

\section{Towards a prediction of earthquakes?}

\subsection{Nature of the problem}

An important effort is carried
out world-wide in the hope that, maybe sometimes in the future, the grail
of useful earthquake prediction will be attained. Among others, the
research comprises  continuous observations of crustal movement and
geodetic surveys, seismic observations, geoelectric-geomagnetic
observations, geochemical and groundwater measurements.
The seismological community has been criticized in the past for
promising results using various prediction techniques (e.g. anomalous seismic
wave propagations,  dilatancy
diffusion, Mogi donuts, pattern recognition algorithms, etc.) that have
not delivered to the expected level. The need for a reassessment of the
physical processes has been recognized and more fundamental studies are
persued on crustal structures in seismogenic zones,
historical earthquakes, active faults, laboratory fracture
experiments, earthquake source processes, etc.

There is even now an opinion that earthquakes could be
inherently unpredictable \cite{Geller}. The argument is that past failures and 
recent theories suggest fundamental obstacles to prediction. It is
then proposed that the emphasis be placed on basic research in
earthquake science, real-time seismic warning systems, and long-term
probabilistic earthquake hazard studies. It is true that
useful predictions are not available at present
and seem hard to get in the near future, but would it not be a little
presumptuous to claim that prediction is impossible? Many
past examples in the development of science
have taught us that unexpected discoveries can modify
completely what was previously considered possible or not. In the 
context of earthquakes, the problem is
made more complex by the societal implications of prediction with, in particular,
the question of where to direct in an optimal way the limited available ressources.

We here focus on the scientific problem and describe a new direction that 
suggests reason for optimism.
Recall that an earthquake is triggered when a mechanical instability occurs and a fracture
(the sudden slip of a fault) appears in a part of the earth's crust. The
earth's crust is in general complex (in composition, strength, faulting) and
groundwater may play an important role. How can one expect to unravel
this complexity and achieve a useful degree of prediction?

\subsection{Large earthquakes}

There is a series of surprising and somewhat controversial studies 
showing that many large earthquakes have been
preceded by an increase in the number of intermediate sized events. The
relation between these intermediate sized events and the subsequent main
event has only recently been recognized because the precursory events occur
over such a large area that they do not fit prior definitions of foreshocks
\cite{JonesMolnar}. In particular, the $11$
earthquakes in California with magnitudes greater than $6.8$ in the last
century are associated with an increase of precursory intermediate magnitude
earthquakes measured in a running time window of five years \cite{Knopoff}. What is
strange about the result is that the precursory pattern occured with
distances of the order of $300$ to $500$ km from the future epicenter, {\it
i.e.} at distances up to ten times larger that the size of the future earthquake rupture. 
Furthermore, the increased intermediate magnitude activity switched
off rapidly after a big earthquake in about half of the cases. This implies
that stress changes due to an earthquake of rupture dimension as small as
$35$ km can influence the stress distribution to distances more than ten
times its size. This result defies usual models.

This observation is not isolated. There is mounting evidence
that the rate of occurrence of intermediate earthquakes increases in the
tens of years preceding a major event. Sykes and Jaume \cite{SykesJaum}
present evidence that the occurrence of events in the range $5.0-5.9$
accelerated in the tens of years preceding the large San Francisco bay area
quakes in 1989, 1906, and 1868 and the Desert Hot Springs earthquake in
1948. Lindh \cite{Lindh} points out references to similar increases in
intermediate seismicity before the large 1857 earthquake in southern
California and before the 1707 Kwanto and the 1923 Tokyo earthquakes in
Japan. More recently, Jones \cite{Jones} has documented a
similar increase in intermediate activity over the past $8$ years in southern
California. This increase in activity is limited
to events in excess of $M = 5.0$; no increase in activity is apparent when
all events $M > 4.0$ are considered. Ellsworth et al. \cite{Ellsworth} also
reported that the increase in activity was limited to events larger
than $M = 5$ prior to the 1989 Loma Prieta earthquake in the San Francisco
Bay area. Bufe and Varnes \cite{BufeVarnes} analysed the increase
in activity which preceded the 1989 Loma Prieta earthquake in the San
Francisco Bay area while Bufe et al \cite{Biufe} document a current increase in
seismicity in several segments of the aleutian arc.

Recently, we have investigated more quantitatively these observations and asked 
what is the law, if any, controlling
the increase of the precursory activity \cite{nous}. Inspired
by our previous considerations of the critical nature of rupture and 
extending it to seismicity,
we have invented a systematic procedure to 
test for the existence of critical behavior and to identify the region 
approaching criticality, based on a comparison of the observed 
cumulative energy (Benioff strain) release and the accelerating 
seismicity predicted by theory. This method has been used to find the 
critical region before all earthquakes along the Californian San Andreas 
system since 1950 with $M \geq 6.5$.  The statistical significance 
of our results was assessed by performing the same procedure on
a large number of randomly generated synthetic catalogs. 
The null hypothesis, that the observed acceleration in all
these earthquakes could result from spurious patterns 
generated by our procedure in purely random catalogs, 
was rejected with $99.5\%$ confidence \cite{nous}. An empirical relation 
between the logarithm of the critical region radius (R) and the
magnitude of the final event (M) was found, such that $\log R \sim 0.5 M$, 
suggesting that the largest probable event in a given region scales 
with the size of the regional fault network.

\subsection{Log-periodicity?}

We must add a third and last touch to the picture, which uses
the concept of discrete scale invariance, its associated complex
exponents and log-periodicity, as discussed above.
In the presence of the frozen
nature of the disorder together with stress amplification effects, we
showed that the critical behavior of rupture is described by complex
exponents, in other words, the measurable physical quantities can exhibit
a power law behavior (real part of the exponents) with superimposed
log-periodic oscillations (due to the imaginary part of the exponents).
Physically, this stems from a spontaneous organization on a fractal fault
system with ``discrete scale invariance''. The practical upshot is that
the log-periodic undulations may help in ``synchronizing'' a better fit to
the data. In the above numerical model, most of the large earthquakes
whose period is of the order of a century can be predicted in this way $4$
years in advance with a precision better than a year. For the real earth,
we do not know yet as several difficulties hinder a practical
implementation, such as the definition of the relevant space-time domain.
A few encouraging results have been obtained but much remains
to test these ideas systematically, especially using the methodology 
presented above to detect the regional domain of critical maturation before
a large earthquake \cite{SS}.

While encouraging and suggestive,  extreme caution should be
exercized before even proposing that this method is useful for predictive
purpose (see \cite{Huangarti,Huangloma} for potential problems and 
\cite{Sobolev,newkobe,Ouillonmine} for positive evidence)
but the theory is beautiful in its self-consistency and, even if
probably inacurate in details, it may provide a useful guideline for the future.

\section{Turbulence and intermittent climate change}

\begin{figure}
\parbox[l]{7.5cm}{
\epsfig{file=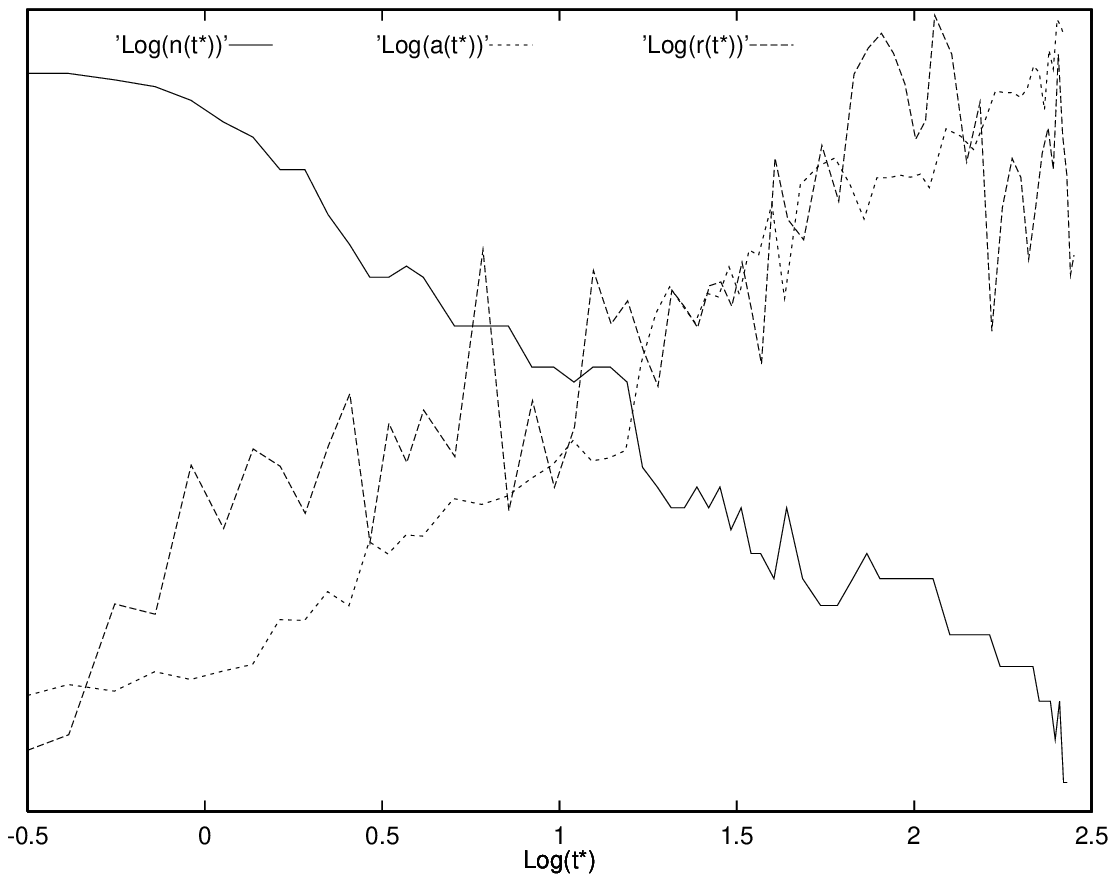,width=8cm}
\caption{\label{rub1} Natural logarithm of the separation, radius and number of vortices as a
function of the natural logarithm of the time
in 2-d experiments of freely decaying turbulence. The vertical scale is different
for each time series ($1$ to $2$ for the logarithm of the separation, $-0.5$ to $0.1$
for the logarithm of the radius and $2$ to $4$ for the logarithm of the number of 
vortices). The approximate linear
dependences in this log-log representation qualify scaling \cite{Weisswill}.
The decorating fluctuations are shown in Fig.\ref{jqmqm} to be genuine
structures characterizing an intermittent dynamics  \cite{Johturb2D}.} }
\hspace{5mm}
\parbox[r]{7.5cm}{
\epsfig{file=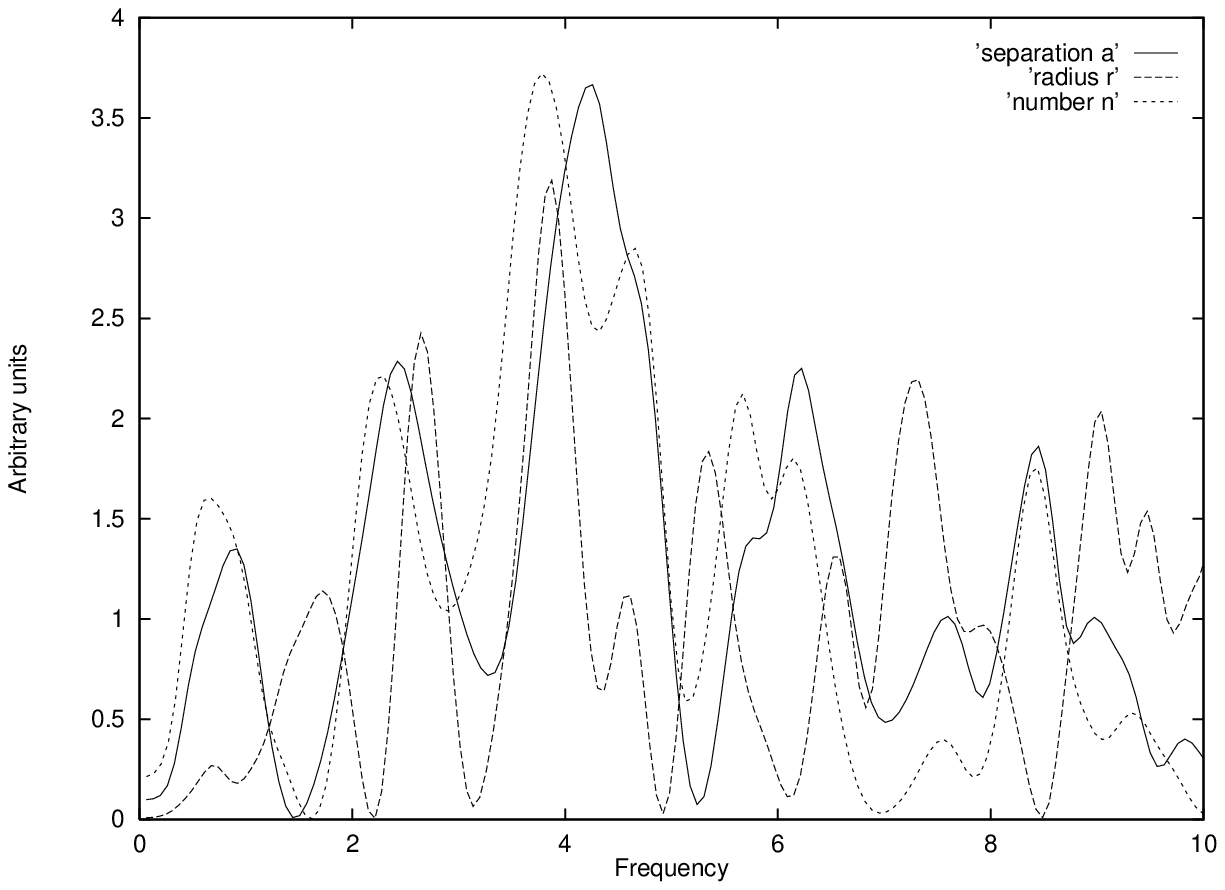,width=8cm}
\caption{\label{jqmqm} Power spectrum of the fluctuations
in the variable $\ln t^*$ of the data shown in Fig.\ref{rub1}. All three
variables (separation, radius and number of vortices) give a peak at the same
log-frequency close
to $4$, corresponding to a prefered scaling ratio of $\lambda \approx 1.3$
 \cite{Johturb2D}.} \vspace{5mm} }
\end{figure}

Turbulence is of special interest, both for its applications in
astrophysics and geophysics and its theoretical properties.
Its main characteristic property is the formation of and
interaction between coherent structures or vortices.
To illustrate the possible existence and importance of specific features with
potential application in geophysical modelling,
we identify a novel signature of turbulent flows in two dimensions \cite{Johturb2D}.
In Figures \ref{rub1} and \ref{jqmqm}, we analyze
freely decaying 2-D turbulence experiments \cite{HanTabel} and document
log-periodic oscillations in the time evolution of
the number of vortices $n(t^*)$, their radius $r(t^*)$
and separation $a(t^*)$, which are the
hallmarks of a discrete hierarchical structure with a prefered scaling ratio.
Physically, this reflects the fact that
the time evolution of the vortices is not smooth but punctuated, leading
to a prefered scale factor found equal roughly to $1.3$.
Fig.\ref{rub1} shows the logarithm $\ln (a(t^*))$ of the separation in the
range
$[0.95:1.96]$, the logarithm $\ln (r(t^*))$ of the radius
in the range $[-0.51:-0.11]$ and the logarithm $\ln (n(t^*))$ of the
number of vortices in the range $[2.1:4.1]$, as a function of the logarithm of rescaled time.
A straight line corresponds to a power law. We see that the predictions of
the usual scaling theory \cite{Weisswill}
are approximately verified but that significant structures are superimposed
on the
linear behavior.  Fig.\ref{jqmqm} shows the spectrum of these structures.
Synthetic tests show that the common peak at a log-frequency around $4$ is
significant \cite{Johturb2D}.

Demonstrating
unambiguously the presence of log-periodicity and thus of 
discrete scale invariance in more general three-dimensional turbulent
time-series would provide an important step towards a direct demonstration of the
Kolmogorov cascade or at least of its hierarchical imprint.
For partial indications of log-periodicity in turbulent data, 
we refer the reader to fig. 5.1 ~p.58 and fig. 8.6~ p.128 of Ref.
\cite{Frisch}, fig.3.16~ p. 76 of Ref.\cite{Arneodo}, fig.1b of Ref.\cite{Tcheou} and 
fig. 2b of Ref. \cite{Castaing}. See \cite{ETCfrisch}
for a proposed mechanisn in which scale invariant equations
that present an instability at finite wavevector $k$ decreasing with the field
amplitude may generate naturally a discrete hierarchy of internal scales.

Relying on the well-known analogies between 2-D and quasi-geostrophic
turbulence \cite{Salmon}, this example illustrates the potential
of our approach for geophysical flows and climate dynamics \cite{GhilChild},
 as well as for abrupt change of weather regimes.
Indeed, considerable attention has been given recently to the
likelihood of fairly sudden, and possibly catastrophic, climate change
\cite{Broecker}.
Climate changes on all time scales \cite{Mitchell}, although no simple
scaling behavior
has been detected in climate time series. This demands
a careful study of records of climate variations with more sophisticated
statistical methods. Interactions among multiple space and time scales are
typical of a climate system that includes the ocean's \cite{McWilliams}
and atmosphere's
turbulent motions \cite{Salmon}.

On the time scale of $10^4-10^6$ years, it was held for about two decades
that climate variations are fairly gradual, and mostly dictated by
secular variations in the Earth's orbit around the Sun; the latter being
quasi-periodic, with a few dominant periodicities between 20,000 and
400,000 years.  More recently, however, it has become clear
that the so-called mid-Pleistocene transition, at about 900,000 years
before the present, involved a fairly sudden jump in the mean global ice
volume and sea-surface temperatures, accompanied by a jump in the
variance of proxy records of these quantities \cite{Maasch}.
A subcritical Hopf bifurcation in a coupled ocean-atmosphere-cryosphere
model has been proposed as an explanation for the suddenness of this
transition (\cite{Ghil94} and references therein).

At the other end of the paleoclimate-variability spectrum, sudden
changes in the ocean's thermohaline circulation have been associated
with the so-called Heinrich \cite{Heinrich} events of rapid excursions in the
North Atlantic's polar front, about 6000-7000 years apart.  These could
arise from threshold phenomena that accompany highly nonlinear
relaxation oscillations in the same coupled model mentioned above \cite{Ghil94}, 
or in variations thereof.

On the time scales of direct interest to a single human life span, a
fairly sharp transition in Northern Hemisphere instrumental
temperatures and other fields has been claimed for the mid-1970s
\cite{Graham}.  This might, in turn, be related to changes in the duration or
frequency of occurence of weather regimes within a single season 
\cite{ChencWallace,Kimoto1,Kimoto2}. The episodic character of weather
patterns and their organization into persistent regimes has been associated
with multiple equilibria \cite{Cherneydevore} or
multiple attractors \cite{LegrasGhil} in the equations governing
large-scale atmospheric flows.

Sudden transitions, {\it i.e.}, transitions that occur over
time intervals that are short compared to what is otherwise the
characteristic time scale of the phenomena of interest, are obviously
harder to predict than the gradual changes on that time scale. We suggest that the
use of methods discussed here could be useful for developing
forecasting models for rapid climatic transition.

\section{Predicting financial crashes?}

Stock market crashes are momentous financial events that are fascinating to 
academics and practitioners alike. Within the efficient markets literature, 
only the revelation of a dramatic piece of information can cause a crash, yet 
in reality even the most thorough {\em post-mortem} analyses are typically 
inconclusive as to what this piece of information might have been. For 
traders, the fear of a crash is a perpetual source of stress, and the onset of the 
event itself always ruins the lives of some of them, not to mention the 
impact on the economy.

A few years ago, we advanced the hypothesis \cite{crash1}-\cite{crash7}
that stock market crashes are caused by the slow buildup 
of powerful ``subterranean forces'' that come together in one critical instant. 
The use of the word ``critical'' is not purely literary here: in mathematical 
terms, complex dynamical systems such as the stock market can go through 
so-called ``critical'' points, defined as the explosion to infinity of a normally 
well-behaved quantity. As a matter of fact, as far as nonlinear dynamic 
systems go, the existence of critical points may be the rule rather than the 
exception. Given the puzzling and violent nature of stock market crashes, it 
is worth investigating whether there could possibly be a link. 

In doing so, we have found three major points. First, it is entirely 
possible to build a dynamic model of the stock market exhibiting well-
defined critical points that lie within the strict confines of rational 
expectations, a landmark of economic theory,
and is also intuitively appealing. We stress the importance of using
the framework of rational expectation in contrast to many other recent attempts.
When you invest your money in the stock market, in general you do not
do it at random but try somehow to optimize your strategy with your limited
amount of information and knowledge. The usual criticism addressed to theories 
abandoning the rational behavior condition is that the universe of
conceivable irrational behavior patterns is much larger than the set
of rational patterns. Thus, it is sometimes claimed that allowing for 
irrationality opens a Pandora's box of ad hoc stories that have little 
out-of-sample predictive powers. To deserve consideration, a theory should be
parsimonious, explain a range of anomalous patterns in different contexts, and
generate new empirical implications.

Second, we find that the mathematical 
properties of a dynamic system going through a critical point are largely 
independent of the specific model posited, much more so in fact than 
``regular'' (non-critical) behavior, therefore our key predictions should be 
relatively robust to model misspecification. 

Third, these predictions are 
strongly borne out in the U.S.~stock market crashes of 1929 and 1987\,: 
indeed it is possible to identify clear signatures of near-critical behavior 
many years before the crashes and use them to ``predict'' (out of sample) the 
date where the system will go critical, which happens to coincide very 
closely with the realized crash date. We also discovered in a systematic testing
procedure a signature of near-critical behavior that culminated in a two weeks 
interval in May 1962 where the stock market declined by $12\%$. The fact that we
``discovered'' the ``slow crash'' of 1962 without prior knowledge of it just
be trying to fit our theory is a reassuring sign about the integrity of the method.
Analyses of more recent data
showed a clear maturation towards a critical instability that can be
tentatively associated to the 
turmoil of the US stock market at the end of october 1997. It may come as 
a surprise that the same theory is applied to epochs so much different 
in terms of speed of communications and connectivity as 1929 and 1997. 
It may be that what our theory addresses is the question\,: 
has human nature changed?

\subsection{The Crash}

A crash happens when a large group 
of agents place sell orders simultaneously. This group of agents must create 
enough of an imbalance in the order book for market makers to be unable to 
absorb the other side without lowering prices substantially. One curious fact 
is that the agents in this group typically do not know each other. They did not 
convene a meeting and decide to provoke a crash. Nor do they take orders 
from a leader. In fact, most of the time, these agents disagree with one 
another, and submit roughly as many buy orders as sell orders (these are all 
the times when a crash {\em does not} happen). The key question is: by what 
mechanism did they suddenly manage to organize a coordinated sell-off? 

We propose the following answer: all the traders in the world are organized 
into a network (of family, friends, colleagues, etc) and they influence each 
other {\em locally} through this network. Specifically, if I am directly 
connected with $k$ nearest neighbors, then there are only two forces that 
influence my opinion: (a) the opinions of these $k$ people; and (b) an 
idiosyncratic signal that I alone receive. Our working assumption here is that 
agents tend to {\em imitate} the opinions of their nearest neighbors, not 
contradict them. It is easy to see that force (a) will tend to create order, while 
force (b) will tend to create disorder. The main story that we are telling in 
here is the fight between order and disorder. As far as asset prices are 
concerned, a crash happens when order wins (everybody has the same 
opinion: selling), and normal times are when disorder wins (buyers and 
sellers disagree with each other and roughly balance each other out). We 
must stress that this is exactly the opposite of the popular characterization of 
crashes as times of chaos.
 
We claim that models of a crash that combine the following 
features:
\begin{enumerate}
\item A system of noise traders who are influenced by their neighbors;
\item Local imitation propagating spontaneously into global cooperation;
\item Global coperation among noise traders causing a crash;
\item Prices related to the properties of this system;
\item System parameters evolving slowly through time;
\end{enumerate}
would display the same characteristics as ours, namely prices following a 
power law in the neighborhood of some critical date, with either a real or 
complex critical exponent. What all models in this class would have in 
common is that the crash is most likely when the locally imitative system 
goes through a {\em critical} point. 

Strictly speaking, these renormalization group equations 
developed to describe the approach to a critical market crash 
are approximations valid only in the 
neighborhood of the critical point. We have proposed a 
more general formula with additional degrees of freedom to better capture 
the behavior away from the critical point. The specific way in which these 
degrees of freedom are  introduced is based on a finer analysis of the 
renormalization group theory that is equivalent to including the next term in 
a systematic expansion around the critical point and introduce a log-periodic
component to the market price behavior.

\subsection{Extended efficiency and systemic instability}

Our main point is that the market anticipates the crash in a
subtle self-organized and cooperative fashion, hence releasing precursory
``fingerprints'' observable in the stock market prices. In other words, this
implies that market prices contain information on impending crashes. 
Our results suggest a weaker form of the ``weak efficient market
hypothesis'' \cite{Fama}, according to which the market prices contain, in
addition to the information generally available to all, subtle informations
formed by the global market that most or all individual traders have not yet
learned to decipher and use. Instead of the usual interpretation of the
efficient market hypothesis in which traders extract and incorporate
consciously (by their action) all informations contained in the market prices,
it may be that the market as a whole can exhibit an ``emergent'' behavior not
shared by any of its constituants. In other words, 
we have in mind the process of
the emergence of intelligent behaviors at a macroscopic scale that
individuals at
the  microscopic scale have no idea of. This process has been discussed in
biology for instance in animal populations such as ant colonies, or in
connection with the emergence of conciousness \cite{Anderson,Holland}. 
The usual efficient market
hypothesis will be recovered in this context when the traders learn how to
extract this novel collective information and act on it.

Most previous models proposed for crashes have pondered the possible mechanisms
to explain the collapse of the price at very short time scales. Here in
contrast, we propose that the underlying cause of the crash must be
searched for years before it in the progressive accelerating ascent of the market
price, the speculative bubble,
 reflecting an increasing built-up of the market cooperativity. From
that point of view, the specific manner by which prices collapsed is not
of real importance since, according to the concept of the critical point, any
small disturbance or process may have triggered the instability, once ripe. The
intrinsic divergence of the sensitivity and the growing instability of the
market
close to a critical point might explain why attempts to unravel the local origin
of the crash have been so diverse. Essentially all would work once the system is
ripe. Our view is that the crash has an endegeneous origin and that exogeneous
shocks only serve as triggering factors. We propose that the origin of the
crash is
much more subtle and is constructed progressively by the market as a whole. In
this sense, this could be termed a systemic instability. This understanding offers
ways to act to mitigate the build-up of conditions favorable to crashes.

A synthesis on the status of this theory
as well as  all available tests on more than 20 crashes
are reported in \cite{QGSJ}.

\section{Predicting human parturition?}

Parturition is the act of giving birth. While not usually considered as
catastrophic, it is arguably the major event in a life (apart from its termination)
and it is interesting that our theoretical approach extends to this situation.
This is not so surprising in view of the commonalities with the previous
examples. 

Can we predict parturition? Notwithstanding
the large number of investigations on the factors that could trigger 
parturition in superior mammals (monkeys and humans), we still do not
have a clear signature in any of the measured variables. This is in contrast 
with the situation for inferior mammals such as cats, cows, etc, for
which the secretion of a specific hormone can be linked unambiguously to 
the triggering of parturition.

Knowledge of precursors and predictors of human parturition would be
important both for our understanding of the controlling mechanisms and for
practical use for detection and diagnostic of various abnormalities of
the birth process. They involve a multitude of genetic, metabolic, nutritional,
hormonal and environmental factors. Present research is however hindered by
the lack of a clear recognized correlation between the time evolution of
these various variables with the initiation of parturition. 

\subsection{Critical theory of parturition}

In collaboration with a team of obstetricians, we have
proposed \cite{Ferre} a coherent logical framework which allows us to rationalize the
various laboratory and clinical observations on the maturation, the
triggering mechanisms of parturition, the existence of various abnormal
patterns as well as the effect of external stimulations of various kinds.
Within the proposed mathematical model, parturition is seen as a ``critical''
instability or phase transition from a state of quietness, characterized by
a weak incoherent activity of the uterus in its various parts as a function
of time (state of activity of many small incoherent intermittent
oscillators), to a state of globally coherent contractions where the uterus
functions as a single macroscopic oscillator leading to the expulsion of the baby. 
Our approach gives a number
of new predictions and suggests a strategy for future research and clinical
studies, which present interesting potentials for improvements in
predicting methods and in describing various prenatal abnormal situations.

We have proposed
to view the occurrence of parturition as an instability,
in which the control parameter
is a maturity parameter (MP), roughly proportional to time, and the
order parameter is
the amplitude of the coherent global uterine activity in the parturition
regime. This idea is summarized by the concept of a so-called
supercritical ``bifurcation''.
This simple view is in apparent contradiction with the extreme complexity
of the fetus-mother system, 
which can be addressed at several levels of descriptions, starting
at the highest level from the mother, the fetus and their coupling through
the placenta. For example, in the mother, the myometrium 
plays an important role in pregnancy, maturation and onset of labor.
It is now well-established that the human myometrium is an heterogeneous
tissue formed of several layers which differ in their embryological origin
and which exhibit quite different histological and pharmalogical
properties. In the uterine corpus, one must distinguish the outer
(longitudinal) and the inner (circular) layers. These two layers composed
mainly of smooth muscle cells are separated by an intermediate layer
containing a large amount of vascular and connective tissues, but poor in
smooth muscle cells. The inner and outer muscle layers have different
patterns of contractility and differ in their response and sensitivity to
contractile and relaxant agents. This is just an example of the complexity
which goes on down to the molecular level, with the action of many
substances providing positive and negative feedbacks evolving as a function
of maturation. The basis of our simple theory relies on many recent works
in a variety of domains (mathematics, hydrodynamics, optics,
chemistry, biology, etc) which have shown that a lot of complex systems
consisting of many nonlinear coupled subsystems or components may
self-organize and exhibit coherent behavior of a macroscopic scale in time
and/or space, in suitable conditions. The Rayleigh-B\'enard fluid
convection experiment is one
of the simplest paradigms for this type of behavior.
The coherent behavior appears generically when the coupling between the
different components becomes strong enough to trigger or synchronize the
initially incoherent subsystems. There are many observations in human
parturition where an increasing ``coupling'' is associated with maturation of
the fetus leading to the cooperative synchronized action of all muscle
fibers of the uterus characteristic of labor. 

\subsection{Predictions}

Perhaps, the most vivid illustration of the increasing coupling as
maturation increases is provided by monitoring the uterine activity, using
standard external tocographic techniques. Away from
term, the muscle contractions during gestation are generally weak and
characterized by local bursts of activity both in time and space.
Increasing uterine activity is observed when the term is approaching,
culminating in a complete modification of behavior where regular globally
coherent contractions reflects the spatial and time coherence of all the
muscles constituting the uterus. The transition between the premature
regime and the parturition regime at maturity is characterized by a
systematic tendency to increasing uterine activity, both in amplitude,
duration of the bursts and spatial extension of the activated uterine
domains. The susceptibility of the fetus-mother system (to influence the
uterine response) to external perturbations or stimulations seems to
increase notably on the approach of parturition, since important
modifications and reactions of the uterus may result from relatively small
stimulii from the mother or fetus.

The main prediction is that, on the
approach to the critical instability, one expects a characteristic increase
of the fluctuations of uterine activity. Other quantities that could be
measured and which are related to the uterine activity are expected
to present a similar behavior.
The cooperative nature of maturation and parturition proposed
here rationalizes the present inability to establish unequivocally
predictive parameters of the biochemical events preceeding myometrical
activity and/or cervical ripening involved in preterm labor. Our theory
suggests a precise experimental methodology in order to obtain an early
diagnosis, essential for the efficient treatment of prematurity, which
still constitutes the major cause of neonatal morbidity and mortality. In
particular, monitoring muscle tremors or vibrations 
 as a function of time of muscle fibers of the uterus would provide
quantitative tests of the theory with respect to the spatio-temporal
build-up of contractile fluctuations. Our theory also correctly accounts
for the observations that external factors affecting the mother such as
heavy work and psychologic stress are able to modify
the maturity of the uterus measured by the progressive modification of the
cervix and more frequent uterine contractions. These external factors, in
addition to produce direct contraction stimulations,
  could also be able to modify the post-maturity
parameter and control the susceptibility of the
fetus-mother system to small influences which can trigger the change from
discordant contrations to concordant contractions of a premature or
post-mature labor. 

We note finally that the whole policy for the
description of risk factors has been based on an implicit and unformalized
hypothesis of a critical transition which is explicited in our
theoretical framework. The
prevention program for preterm deliveries \cite{Papiernik1,Papiernik2} was also based
on the hypothesis of such a critical transition and the understanding that
a small reduction of a triggering factor could be enough to prevent the
uterus from beginning its critical phase of activity. The high
susceptibility of the fetus-mother system to various factors is also at the
origin of the fact that the conventional system of calculation of the risk
factors does not explain the real success of the prevention which has been
observed \cite{Papiernik1,Papiernik2}. Effectively applied in France, our
system, which is based on this idea of a critical transition, was able to
reduce significantly the rate of preterm births for all french women
measured on Haguenau population of pregnant women from 1971 to 1982, 
or on randomized samples of all french births.

\section{Summary}

We have proposed that catastrophes, as they occur in various disciplines,
have similarities both in the 
failure of standard models and the way that systems evolve towards them.
We have presented a non-traditional general methodology for the
scientific predictions of catastrophic events, 
based on the concepts and techniques of statistical and nonlinear physics.
This approach provides a third line of attack bridging across
the two standard strategies of analytical theory and brute force
numerical simulations. It has been successfully applied to problems as
varied as failures of engineering structures, stock market crashes and human
parturition, with potential for earthquakes. 

The review has emphasized the statistical physics point of view, 
stressing the cascade across scales and the interactions at multiple scales.
Recently, a simple two-dimensional dynamical system has been 
introduced \cite{IdeSor} which reaches
a singularity (critical point) in finite time decorated by accelerating oscillations due to
the interplay between nonlinear positive feedback and 
reversal in the inertia. This dynamical system has been shown
to provide a fundamental equation for the 
dynamics of (1) stock market prices going to a crash in the presence of 
nonlinear trend-followers and nonlinear value investors, (2) 
the world human population with a competition between a
population-dependent growth rate and a nonlinear 
dependence on a finite carrying capacity and
(3) the failure of a material
subject to a time-varying stress with a competition between positive
geometrical feedback on the damage variable and nonlinear healing.
The rich fractal scaling properties of the dynamics have been traced back to
the self-similar spiral structure in phase space
unfolding around an unstable spiral point at the origin. We expect 
exciting new developments in this field using a combination of 
statistical physics and dynamical system theory.

Let us finally mention possible extensions of this approach for 
future research on the prediction of societal breakdowns, terrorism, large scale 
epidemics, and of the vulnerability of civilisations \cite{endera}.

{\bf Acknowledgments}: I am grateful to my co-authors listed in the bibliography
for stimulating and enriching collaborations and to D. Turcotte for
a careful reading of this manuscript. This work was partially supported by 
NSF-DMR99-71475 and the James S. Mc Donnell Foundation 21st century scientist award/studying
complex system.

\clearpage

\clearpage

\clearpage

\begin{figure}
\parbox[l]{7.5cm}{
\epsfig{file=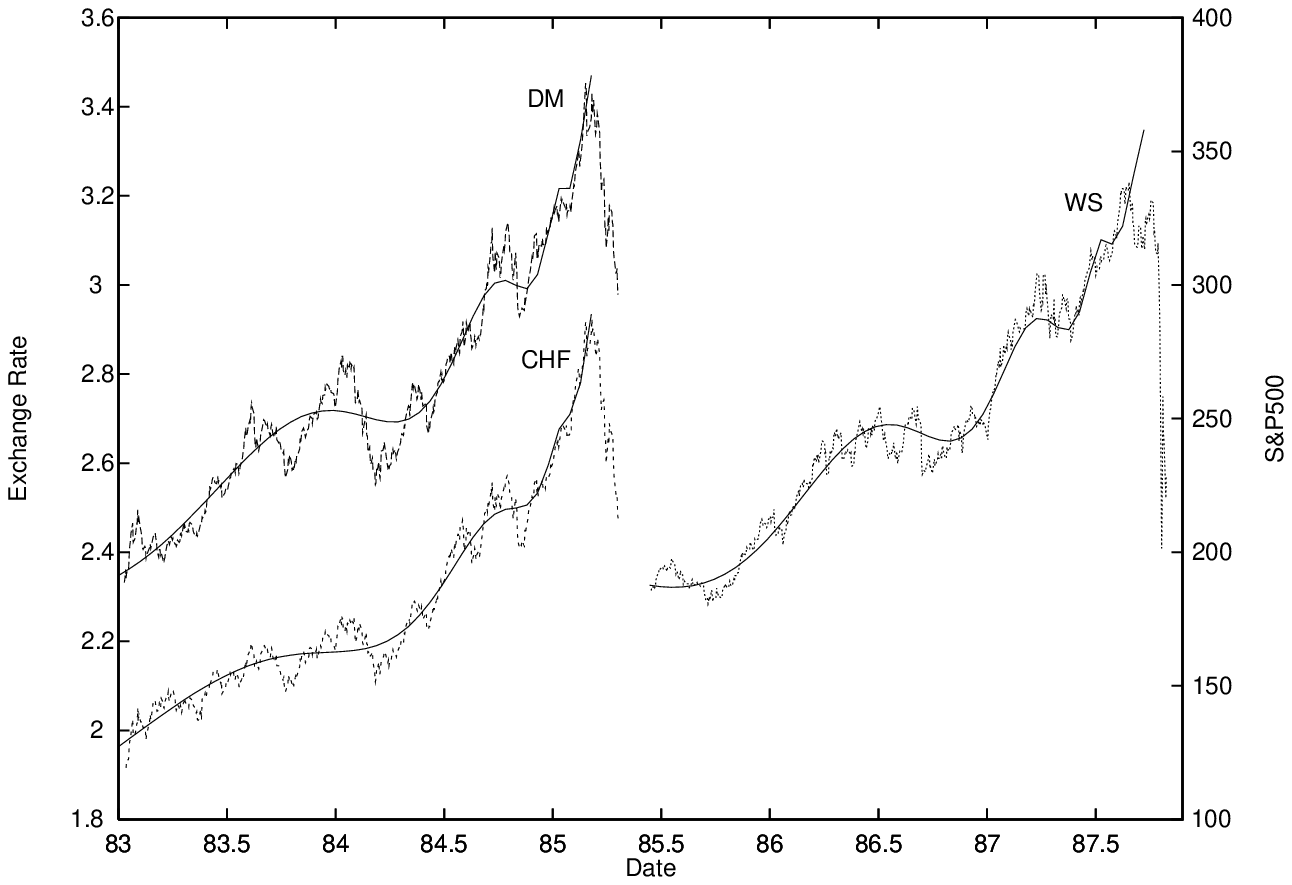,width=8cm}
\caption{\label{crashaaa} The S\&P 500 US index prior to the October
1987 crash on Wall Street and the US \$ against DEM and CHF prior to the
collapse in mid-1985. The continuous lines are fits with a power law modified
by log-periodic oscillations. See \cite{crash1}-\cite{crash7}.} }
\hspace{5mm}
\parbox[l]{7.5cm}{
\epsfig{file=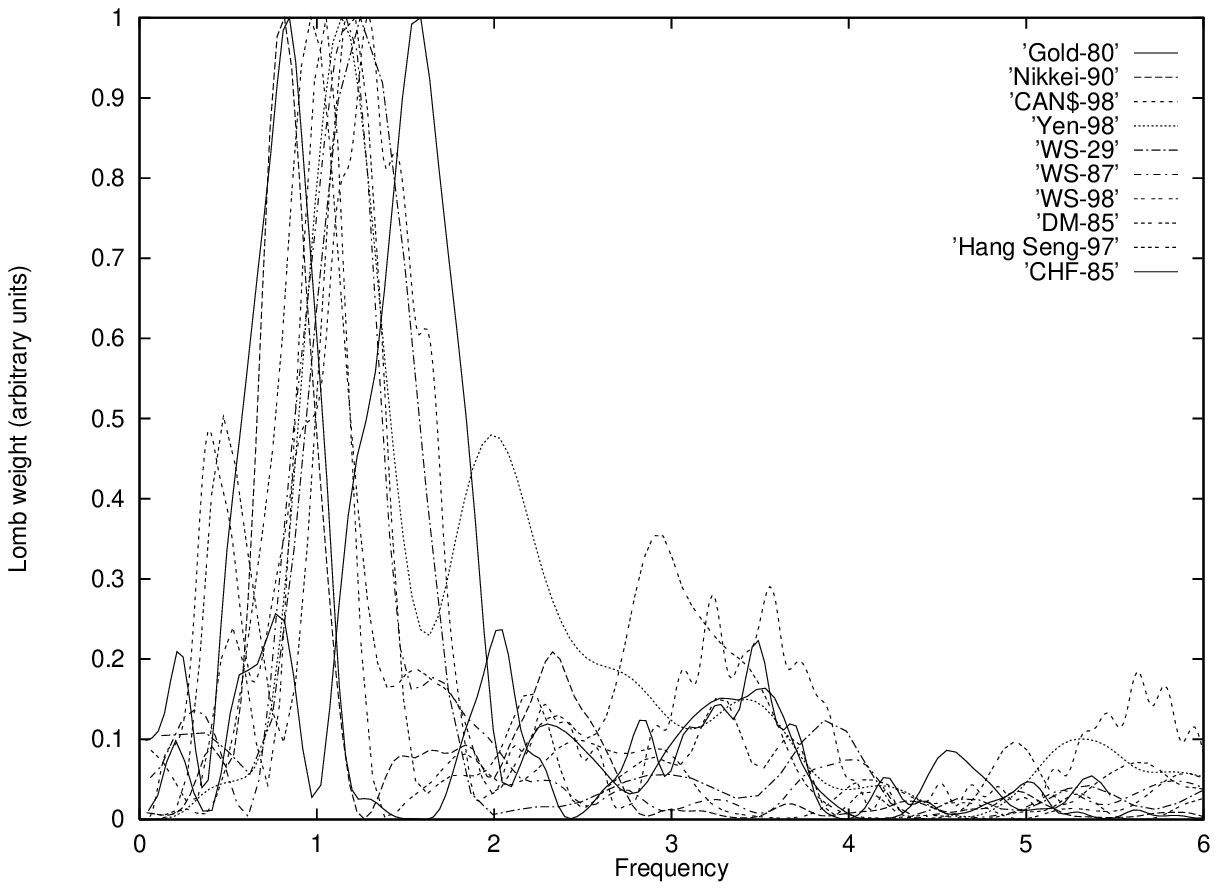,width=8cm}
\caption{\label{lombcrahs} Power spectrum (Lomb periodogram) for 10 major
events in this century, showing a very robust and intertemporal pattern associated with
discrete scale invariance with a well-defined
prefered scaling ratio $\lambda$. Such structures can be used in
forecasting mode with good skills. See \cite{crash1}-\cite{crash7}.}\vspace{1cm}}
\end{figure}

\end{document}